\documentclass{statsoc}

\usepackage{geometry}
\usepackage[utf8]{inputenc}
\usepackage{amsfonts, amsmath, amssymb}
\usepackage{graphics,graphicx}
\usepackage{natbib}
\usepackage{subfigure}
\usepackage{multirow}
\usepackage[normalem]{ulem}
\usepackage{tikz,url}

\newcommand{\bfX}{{\bf X}}
\newcommand{\bfbeta}{\mbox{\boldmath $\beta$}}

\DeclareGraphicsExtensions{.pdf,.png,.jpg}
\title[Abundance model for misuse prevalence]{An integrated abundance model for estimating county-level prevalence of opioid misuse in Ohio}

\author{Staci A. Hepler\textsuperscript{\footnotemark[1]}}
\address{Department of Mathematics and Statistics, Wake Forest University, Winston-Salem, USA}
\email{heplersa@wfu.edu}
\author{David M. Kline\textsuperscript{\footnotemark[1]}}
\address{Department of Biostatistics and Data Science, Wake Forest School of Medicine, Winston-Salem, USA}
\email{dkline@wakehealth.edu}
\author{Andrea Bonny}
\address{Division of Adolescent Medicine, Nationwide Children's Hospital, Department of Pediatrics, The Ohio State University, Columbus, USA}
\author{Erin McKnight}
\address{Division of Adolescent Medicine, Nationwide Children's Hospital, Department of Pediatrics, The Ohio State University, Columbus, USA}
\author[Hepler {\it et al.}]{Lance A. Waller}
\address{Department of Biostatistics and Bioinformatics, Emory University, Atlanta, USA}

\begin{document}

\maketitle

\footnotetext[1]{Authors contributed equally}

\begin{abstract}
Opioid misuse is a national epidemic and a significant drug related threat to the United States. While the scale of the problem is undeniable, estimates of the local prevalence of opioid misuse are lacking, despite their importance to policy-making and resource allocation. This is due, in part, to the challenge of directly measuring opioid misuse at a local level. In this paper, we develop a Bayesian hierarchical spatio-temporal abundance model that integrates indirect county-level data on opioid-related outcomes with state-level survey estimates on prevalence of opioid misuse to estimate the latent county-level prevalence and counts of people who misuse opioids. A simulation study shows that our integrated model accurately recovers the latent counts and prevalence. We apply our model to county-level surveillance data on opioid overdose deaths and treatment admissions from the state of Ohio. Our proposed framework can be applied to other applications of small area estimation for hard to reach populations, which is a common occurrence with many health conditions such as those related to illicit behaviors. 
\end{abstract}


\maketitle

\section{Introduction}

The opioid epidemic in the United States is a public health crisis \citep{WH2011,DEA} associated with unprecedented morbidity and mortality \citep{Rudd2016,Zibbell2015}. In the 12 month period ending in April 2021, drug overdose deaths in the United States exceeded 100,000, representing a 30\% increase from the prior year \citep{CDCweb2021b}. The opioid epidemic has been particularly severe in Ohio.  In 2019, Ohio had the fourth highest overdose death rate of 38.3 per 100,000 which was nearly double the national rate of 21.6 per 100,000 \citep{CDCweb2021}. In addition to the epidemic of overdose death, opioid misuse puts Ohio at risk of epidemic levels of Hepatitis C and HIV \citep{Lerner2019}.  Knowledge of local prevalence of people who misuse opioids (PWMO) is imperative to quantifying the magnitude of the opioid epidemic. However, this quantity is an unobservable, dynamic subset of the population, and the lack of local estimates of PWMO is a significant barrier to the public health response to the opioid crisis \citep{Schuler2020}.

Despite its importance for guiding a public health response, surveillance of behaviorally-linked conditions, like substance use, is challenging.  These conditions tend to vary at the local level and rely on self-reported information, which may not be reliable when dealing with an illicit behavior.  This limits the utility of large surveys because they are not typically designed to provide local estimates and are reliant on self-reporting \citep{Palamar2016}.  While respondent driven sampling designs \citep{Handcock2014,Crawford2018} have been implemented to generate estimates specific to local areas for hard to reach populations, they are typically cross-sectional and would be logistically difficult to implement across larger areas of interest. Thus, there is a great need for local, longitudinal, population-level estimates that cover an entire area of administrative interest, like a state.

In Ohio, the local spatial unit of interest is the county, as that most closely corresponds to the structure of local health districts which form the basis for the allocation of state resources. We lack direct information on the prevalence of PWMO in all of Ohio's counties. However, we do have indirect information on this quantity in the form of surveillance data collected at the county level.  Specifically, we have access to counts of overdose deaths and treatment admissions annually for each county. However, neither surveillance outcome is a perfect reflection of PWMO because there are unique selection processes associated with capturing an individual in either observed outcome that are likely heterogeneous across space and time.  For example, overdose death rates are likely related to supply of fentanyl in the local drug market \citep{Pardo2019}.  Likewise, treatment admissions are related to local access to care. This heterogeneity prevents us from simply assuming that there is a perfect correlation between the surveillance outcomes and underlying unobservable prevalence of PWMO. Instead, we must explicitly account for these selection processes.

This is structurally similar to the problem faced in ecological applications of estimating population abundance with unmarked observations and unknown detection (i.e., selection) probabilities.  Briefly, abundance models use hierarchical models to estimate the total number of individuals in a community when only a proportion can be observed at a given place and time \citep{Royle2004,Kery2010}. Conceptually, this is also similar to capture-recapture approaches, which have been applied to drug use and HIV in the past \citep{Jones2016,Bao2015,Barocas2018,Min2020}, but does not require identification and tagging of individuals. For our problem, the county-level surveillance outcomes each reflect a proportion of the true population of PWMO in the county whom we are able to observe or detect. Since we lack individual identifying information, we cannot use a capture-recapture approach and instead will consider an extension of abundance modeling to estimate the count and prevalence of PWMO.

One documented challenge with abundance models is the ability to identify intercept parameters for both the latent abundance process and the detection process, which impedes accurate estimates of absolute abundance. \citet{Royle2004} developed an N-mixture model to estimate population size from count data, but identifiability under this approach requires replicated observations of a closed, or unchanging, population. \citet{Solymos2012} suggested replication is unnecessary provided the number of spatial locations is large and there is at least one continuous covariate that is uniquely related to each distinct process. However, \citet{Knape2015} found that estimates of absolute abundance from an N-mixture model without replication were sensitive to model assumptions on the detection probabilities. 


For our application, we lack replication. That is, we do not have repeated observations of a stable population to allow us to uniquely identify both the intercept in the model for the unobserved population in the community and the intercept in the model for the partial proportion of the population observed. Facing a similar problem, \citet{Stoner2019} used informative prior distributions to identify a hierarchical model for under-reporting using a single observed surveillance outcome.  While this is a reasonable solution to the problem of non-identifiability when such prior information exists, results will be dependent on the choice of prior \citep{Neath1997}. Instead, we integrate multiple sources of data at the county and state levels to identify our model. This is the notion behind integrated population models, which are increasingly popular ecological models used to analyze population size. By jointly modeling multiple data sources that share some underlying latent process, these models can allow estimation of parameters which would be non-identifiable in a single outcome model \citep{Besbeas2002, Schaub2011}. In our setting, the models for the county-level surveillance outcomes of overdose death and treatment counts both relate to the latent number of PWMO. Incorporating multiple county-level surveillance outcomes creates ``pseudo-replication'' because they show aggregate subsets of individuals who were detected, providing multiple partial views of the population of interest. We also incorporate state-level survey estimates of PWMO to inform the overall statewide prevalence, which, in turn, provides information about the range of the detection probabilities required for identification of the model \citep{Knape2015}. Thus, we use indirect count data on county-level surveillance outcomes and an integrated abundance model framework to inform county-level estimates of PWMO. This can also be viewed as an approach for using the county-level surveillance data to inform small area model-based estimates that downscale the state-level survey data to obtain county-level estimates on PWMO.


In this paper, we develop an extension to the abundance modeling framework that integrates multiple aggregate data sources at different spatial supports to estimate county-level prevalence of PWMO. We define PWMO as those who have misused opioids in the past year and are at risk of experiencing overdose death or treatment admission for opioid misuse. In addition to prevalence and relative risks, we estimate county-level counts of PWMO, which may ultimately be more useful for per capita resource allocation. We will illustrate through simulation studies that our model accurately estimates the quantities of interest and is a drastic improvement over baseline approaches ignoring spatial heterogeneity. We will also illustrate the benefits of including multiple surveillance outcomes.  We then apply our methodology to data from Ohio to estimate annual county-level prevalence of PWMO over a 13 year period.  By doing so, we provide a coherent framework for integrating multiple sources of data to estimate an unobservable quantity of critical importance for public health policy and resource allocation.

The rest of the paper is organized as follows.  We describe the available data in Section \ref{se:data}.  The modeling framework is detailed in Section \ref{se:model}. We present the design and results of our simulation study in Section \ref{se:simulation}. In Section \ref{se:app}, we describe the results for the application to the data from Ohio.  We discuss the implications of our findings in Section \ref{se:discuss}.

\section{Data}
\label{se:data}

Since we lack direct county-level data on PWMO, our primary data sources will be annual county-level counts of overdose deaths and treatment admissions for each of Ohio's 88 counties from 2007-2019, the most recent available year.  Overdose death data are publicly available from the Ohio Public Health Data Warehouse \citep{OPHDW}.  Deaths are indexed to the county of residence of the decedent and are counted if the death certificate mentions poisoning from any opioid.  Annual county-level treatment admission counts were obtained through a data use agreement with the Ohio Department of Mental Health and Addiction Services.  Treatment admissions are indexed to the patient's county of residence and capture any residential, intensive outpatient, or outpatient treatment for opioid misuse. Data were provided broken down into two age groups (adolescents and adults) but will only be considered in total for this study.  State policy requires counts between 1 and 9 to be suppressed, causing some counties to have censored counts.  Population data used for calculating rates are estimates from the National Center for Health Statistics and were also obtained from the Ohio Public Health Data Warehouse \citep{OPHDW}.

Our model also incorporates state-level survey estimates of the prevalence of PWMO from the National Survey on Drug Use and Health (NSDUH) \citep{SAMHSAsurvey,SAMHSArdas}. We obtain state-level estimates for past year nonmedical opioid use for surveys prior to 2015 and past year opioid misuse after 2015. The language of the survey question was updated in 2015, but we assume the same underlying construct is addressed over time. The survey data for Ohio are estimates of multi-year averages of the statewide prevalence of misuse. The supplementary material contains the multi-year estimates of the statewide prevalence along with the standard errors. Note that the county-level data are from the years 2007 ($t=1$) to 2019 ($T=13$), but we incorporate survey information from 2003 ($t=-3$) to 2019 into the model. The survey estimates are shown in Supplemental Table 1.

We utilize county-level estimates of sociodemographic characteristics from the American Community Survey (ACS) using the R package \texttt{tidycensus}.  Variables obtained included poverty rate, unemployment rate, the proportion with at least a high school degree, and the proportion on food stamps. ACS estimates of the 5-year averages are available for all $n=88$ counties in Ohio from 2009 - 2019. ACS estimates for the individual years are available for $38$ of the $88$ counties. We also acquired publicly available data on opioid prescribing rates from the Ohio Automated RX Reporting System (OARRS), which was available from 2010-2019.  In addition, we compiled county-level indicators of health professional shortage areas (HPSA) and medically underserved areas (MU) from the Health Resources and Services Administration, high intensity drug trafficking areas (HIDTA) from the Drug Enforcement Agency, and metropolitan statistical areas (MSA) from the United States Census Bureau. We also created an indicator of whether an interstate highway passed through each county to reflect transportation networks. 

\section{Model} 
\label{se:model}

Let $Y_{it}^{(k)}$ be the count of PWMO who experience outcome $k=1,...,K$ in county $i=1,...,n$ during year $t=1,...,T$. Note that in our application, $K=2$ and $k=1, 2$ refer to treatment admissions and overdose deaths, respectively. We also observe state-level survey information, denoted $S_{a:b}$, regarding the estimated statewide average prevalence of opioid misuse for the multi-year time period from year $a$ to year $b$ (inclusive). We are ultimately interested in estimating the latent number of PWMO in county $i$ during year $t$, $N_{it}$, and the relative risk of misuse, $\lambda_{it}$. As in \cite{Berliner1996}, we use a three-stage Bayesian hierarchical model to relate the observed data to the latent processes of interest. The data, process, and prior layers of the model are defined in the following subsections. 

\subsection{Data Model}

\subsubsection{County-level Surveillance Data}
We start by specifying a model for the observed county-level surveillance outcomes. Assume
\begin{equation}
\label{eq:data}
    Y_{it}^{(k)}|N_{it},p_{it}^{(k)} \stackrel{ind}{\sim} \text{Binomial}\left(N_{it},p_{it}^{(k)}\right),
\end{equation}
where 
\begin{equation*}
    \text{logit}\left(p_{it}^{(k)}\right)=\mu_{t}^{(k)}+\bfX_{it}^{(k)} \bfbeta^{(k)}+ f^{(k)}_{it}+\epsilon_{it}^{(k)}.
\end{equation*}
In the logistic regression model for each outcome $k$, $\mu_{t}^{(k)}$ is a time varying intercept, $\bfX_{it}^{(k)}$ is a vector of centered covariates, $\bfbeta^{(k)}$ is a vector of regression coefficients, $f^{(k)}_{it}$ is a spatio-temporal random effect, and  $\epsilon_{it}^{(k)} \overset{iid}{\sim}N(0,\sigma^2_k)$ accounts for additional unexplained heterogeneity.  We assume that each surveillance outcome is conditionally independent given the underlying true number of PWMO. That is, we assume marginal dependence between the counts within a county is attributable to the number of PWMO in that county. For our application, the design matrix for treatment rate $\bfX^{(1)}$ contains indicator variables that identify the counties that are classified as HPSA and MU. The design matrix for death rate $\bfX^{(2)}$ contains indicator variables for whether or not the county contains an interstate, whether the county is a HIDTA, and whether the county belongs to a MSA. 

As mentioned in Section \ref{se:data}, the age-group specific treatment counts are suppressed if they are between 1 and 9.  However, we can incorporate the knowledge that suppressed counts are within that interval into our model by adapting the approach of \citet{Famoye2004} for interval censoring.  There are three cases of censoring that we encounter here, denoted by the indicator:
\begin{align}
    c_{it}=\begin{cases}
0 & \text{total count observed}\\
1 & \text{adolescent count censored}\\
2 & \text{adolescent and adult counts censored}.
\end{cases}
\end{align}
For $c_{it}=0$, we observe both the adult and adolescent counts and so the total treatment count is their sum.  For $c_{it}=1$, we observe the adult count, but the adolescent count is censored so we know that the total count must be between the adult count plus 1 and the adult count plus 9.  Finally, for $c_{it}=2$, both counts are censored so we know the total count must be between 2 and 18. Let $Y^{(1)}_{it0}$ be the adolescent treatment count and $Y^{(1)}_{it1}$ be the adult treatment count.  Then, we have
\begin{align}
    Y^{(1)}_{it}=\begin{cases}
    Y^{(1)}_{it0}+Y^{(1)}_{it1} & c_{it}=0\\
    Y^{(1)}_{it1} & c_{it}=1\\
    0 & c_{it}=2,\\
    \end{cases}
\end{align}
and the likelihood from Equation \ref{eq:data} becomes
\begin{align}
    & L(\mathbf{p}^{(1)}, \mathbf{p}^{(2)}, \textbf{N} | \textbf{Y}^{(1)}, \textbf{Y}^{(2)}) = \left[\prod_{t=1}^{13} \prod_{i=1}^{88} f\left(Y_{it}^{(2)} | p_{it}^{(2)}, N_{it}\right) \right] \times  \nonumber \\ 
    & \times \left[\prod_{t=1}^{13} \prod_{i=1}^{88}  \left[f\left(Y_{it}^{(1)}|p_{it}^{(1)}, N_{it}\right)\right]^{I(c_{it}=0)}
    \left[F\left(Y_{it}^{(1)}+9|p_{it}^{(1)}, N_{it}\right)-F\left(Y_{it}^{(1)}|p_{it}^{(1)}, N_{it}\right)\right]^{I(c_{it}=1)} \right. \\
    & \hspace{.2in}  \times \left. \left[F\left(18|p^{(1)}_{it}, N_{it}\right)-F\left(1|p^{(1)}_{it}, N_{it}\right)\right]^{I(c_{it}=2)} \right] \nonumber
\end{align}
where $F\left(\cdot|p^{(1)}_{it}, N_{it}\right)$ is the cumulative distribution function and $f\left(\cdot|p^{(1)}_{it}, N_{it}\right)$ is the probability mass function of the binomial distribution with population size $N_{it}$ and rate $p^{(1)}_{it}$, and $f\left(\cdot|p^{(2)}_{it}, N_{it}\right)$ is the probability mass function of the binomial distribution with population size $N_{it}$ and rate $p^{(2)}_{it}$.

\subsubsection{State-level Survey Data}
\label{sec:survey}

For the survey information regarding overall statewide prevalence of misuse between years $a$ and $b$, given by $S_{a:b}$, we assume a normal distribution truncated to $(0, 1)$. The standard error $\hat{se}_{a:b}$ is estimated  from the survey and assumed to be known. We assume a truncated normal distribution to enable the direct incorporation of the mean and standard error reported from the survey. In addition, we let $\mu_t$ denote the true latent statewide prevalence of misuse in year $t$ and assume this is linear over time such that $\mu_t = \beta_0^{\mu} + \beta_1^{\mu} t$. Note that this formulation requires survey data to be observed for at least two time periods. More specifically, we assume
\begin{align}
    S_{a:b}|\beta_0^{\mu},\beta_1^{\mu} \sim N_{(0,1)}\left( \frac{1}{b-a+1}\sum_{t=a}^b \mu_t, \hat{se}_{a:b}^2 \right),
    \label{eq:surveymulti}
\end{align} 
so that the mean of $S_{a:b}$ is the mean of the true statewide prevalence during the time period from a to b. The linearity assumption for $\mu_t$ implies the mean function is $$ \frac{1}{b-a+1}\sum_{t=a}^b \mu_t = \beta_0^{\mu} + \beta_1^{\mu} \frac{1}{b-a+1}\sum_{t=a}^b t = \beta_0^{\mu} + \beta_1^{\mu} \frac{b^2+b-a^2+a}{2(b-a+1)}.$$ 
We assume survey estimates are independent of one another, conditional on the true statewide rate of misuse.

\subsection{Process Model}


We are primarily interested in estimating the latent number of PWMO, $N_{it}$, and the relative risk of misuse, $\lambda_{it}$.  We assume a non-canonical, spatial rates-like parameterization \citep{Cressie2005,Cressie2011}:
\begin{align*}
    N_{it}|\lambda_{it} \stackrel{ind}{\sim} \text{Binomial}(P_{it},\mu_t \lambda_{it}),
\end{align*}
where $P_{it}$ is the known population of county $i$ during year $t$.  Let $\lambda_{it}$ represent the relative risk of misuse in county $i$ during year $t$ compared to a statewide average prevalence, $\mu_t$, subject to the constraint $0 < \mu_t \lambda_{it} < 1$.  Recall, $\mu_t$ is assumed to be linear across time and is informed by survey data, as described in Section \ref{sec:survey}. 
We use a non-standard parameterization because the survey data reflects a state average which is not equivalent to the intercept of a standard logistic regression. Note that we could also consider a Poisson model for the latent counts. The large population size and small rates of misuse imply these models will yield near identical results. We chose a binomial model to explicitly allow for the known population size, $P_{it}$. 

Assume $\log(\lambda_{it})= \mathbf{W}_{it} \boldsymbol{\gamma} + u_{it}+v_{it}$, where 
$\mathbf{W}_{t}$ is a design matrix for time $t$ containing centered covariates without an intercept, $u_{it}$ is a spatio-temporal random effect, and $v_{it} \overset{iid}{\sim} N(0,\sigma^2_v)$. For the spatio-temporal random effect $u_{it}$, we assume an intrinsic conditional autoregressive (ICAR) model with an autoregressive of order one (AR(1)) temporal trend \citep{Besag1974}. More specifically, this model assumes for $t=1$
\begin{equation}
    u_{it} | \mathbf{u}_{-i,t},\tau^2_u  \sim N \left( \frac{1}{w_{i+}} \sum_{j} w_{ij} u_{jt}, \frac{\tau^2_u}{w_{i+}} \right),
    \label{eq:uICARt1}
\end{equation} 
and for $t=2,...,T$
\begin{equation}
    u_{it} | \mathbf{u}_{-i,t}, u_{i,t-1},\tau^2_u,\phi_u  \sim N \left( \phi_u u_{i,t-1} + \frac{1}{w_{i+}} \sum_{j} w_{ij} (u_{jt}-\phi_u u_{j,t-1}), \frac{\tau^2_u}{w_{i+}} \right),
    \label{eq:uICARt2}
\end{equation} 
where $\mathbf{u}_{-i,t} = \{u_{jt} : j \neq i\}$, $w_{ij}$ is an indicator that counties $i$ and $j$ share a border, and $w_{i+}=\sum_j w_{ij}$ is the total number of neighbors for county $i$. Let $\mathbf{u}_t = \left(u_{1t},...,u_{nt} \right)'$ denote the vector of random effects during year $t$. The intrinsic models specified by equations \eqref{eq:uICARt1} and \eqref{eq:uICARt2} yield joint distributions with probability density functions
\begin{equation*}
    \begin{aligned}
         \left[\mathbf{u}_1 | \tau^2_u\right] &\propto \exp \left( -\frac{1}{2 \tau^2_u} \mathbf{u}'_1(\mathbf{H}-\mathbf{A})\mathbf{u}_1 \right)  \text{ for } t=1 \\
         \left[\mathbf{u}_t | \mathbf{u}_{t-1}, \tau^2_u, \phi_u \right] &\propto \exp \left( -\frac{1}{2 \tau^2_u} \left(\mathbf{u}_t-\phi_u \mathbf{u}_{t-1}\right)'(\mathbf{H}-\mathbf{A})\left(\mathbf{u}_t-\phi_u \mathbf{u}_{t-1}\right) \right)  \text{ for } t=2,...,T,
    \end{aligned}
\end{equation*}
where $\mathbf{A}$ is the adjacency matrix whose $(i,j)$th element is $w_{ij}$ and $\mathbf{H}$ is a diagonal matrix with $(i,i)$th element $w_{i+}$.
The above are not valid probability densities since the precision $\mathbf{H}-\mathbf{A}$ is less than full rank. However, the ICAR model is a valid process level model provided a centering constraint, $\sum_i u_{it}=0$ for all $t, k$, is enforced \citep{Banerjee2004}. 


In this application, we chose to include standardized county-level covariate information on poverty rate, unemployment rate, the percentage with at least a high school degree, the percentage on food stamps, and the opioid prescribing rate per capita as covariates for relative risk of misuse. The prescribing rate data from OARRS is only available from 2010 to 2019. Thus, this variable is only included in $\mathbf{W}_t$ for $t=4,...,T$.

As discussed in Section 1, it is well known that the single visit abundance model suffers from non-identifiability of intercept parameters. To see this more clearly, one can show that in the single outcome ($K=1$) case, integrating out the latent count $N_{it}$ yields the marginal distribution $Y_{it}^{(k)}|p_{it}^{(k)}, \mu_t, \lambda_{it} \sim \text{Binomial}(P_{it}, \mu_t \lambda_{it} p_{it})$. The product form for the rate implies these quantities, and in particular the intercept parameters $\mu_t$ and $\mu_t^{(k)}$, cannot be individually identified without additional information. However, our model's integration of multiple data sources eliminates this issue. In particular, the survey data directly informs $\mu_t$. Also note that by jointly modeling $K>1$ outcomes, the joint marginal distribution of the outcomes after summing over the possible values of $N_{it}$ does not take the same simple marginal form that results with the single outcome setting. By proposing an integrated model that jointly models multiple data sources, we are able to identify the primary quantities of interest, $N_{it}$ and $\lambda_{it}$. Achieving identifiability by integrating multiple data sources is discussed in \cite{Besbeas2002} and Section 9.2 of \cite{Cole2020}.


The models for the surveillance outcomes depend on spatio-temporal random effects $f_{it}^{(k)}$. These are specified similarly to the spatio-temporal random effects in the process-level model for the relative risk of misuse. In particular, for each outcome $k$, we assume an ICAR model with an AR(1) temporal trend analogous to \eqref{eq:uICARt1}-\eqref{eq:uICARt2} with conditional variance parameter $\tau^2_k$ and temporal autocorrelation parameter $\phi_k$. We assume these random effects are independent across the $k$ outcomes. In more traditional models, we would expect these rates to be correlated within a county because they share a common underlying process, the latent number of PWMO in that location. However, we explicitly account for that process as the outcome models are specified conditional on the latent number of PWMO so that these random effects capture residual spatial variability in the outcome-specific rates of opioid overdose death and treatment admissions. We a priori do not believe these conditional rates would follow similar spatial trends. If instead we were modeling outcomes that were believed to be correlated conditional on $N_{it}$, this model can easily be generalized to a multivariate conditional autoregressive model.

\subsection{ACS Covariate Model}

Our process model relates the relative risk of misuse, $\lambda_{it}$, to centered covariates from the ACS in the design matrix ${\bf{W}}_t$.  However, only 5-year average estimates are available for all 88 counties, which leads to temporal misalignment with the annual estimates of interest. To account for this and the additional uncertainty from using multi-year average estimates, we add an additional layer to the data and process stages of the model, similar to the work of \citet{Bradley2015}, to account for uncertainty in the latent yearly county-level values of these variables. This additional layer leverages the annual estimates that are available in 38 counties to inform the latent annual value in counties where only 5 year average estimates are available.

Let $\hat{\omega}_{it}^{(5)}$ and $\hat{\omega}_{it}^{(1)}$ denote the $5$-year and $1$-year estimates from the ACS for one of the variables of interest with standard errors $\hat{\sigma}_{(5)}$ and $\hat{\sigma}_{(1)}$, respectively.  Let $\omega_{it}$ denote the true latent value of this variable in county $i$ during year $t$. The columns of the design matrix $\mathbf{W}_t$ corresponding to the ACS variables contain standardized variables and are thus functions of these latent $\omega_{it}$. For $t=3,...,T$ and each of the ACS variables included, we assume
\begin{equation}
    \begin{aligned}
        \hat{\omega}_{it}^{(5)} &\sim N_{(0,100)} \left( \frac{1}{5} \sum_{\ell=t-4}^t \omega_{i \ell}, \hat{\sigma}_{(5)}^2 \right)\\
        \hat{\omega}_{it}^{(1)} &\sim  N_{(0,100)} \left(  \omega_{it}, \hat{\sigma}_{(1)}^2 \right),
    \end{aligned}
\end{equation}
where $N_{(0,100)}$ denotes the normal distribution truncated to $(0,100)$ since our variables of interest are recorded as percentages. Observe that even though we only include ACS estimates from 2009 - 2019, we infer latent values of the yearly variables from 2005 - 2019.

In the process level of the model, we assume the latent yearly percentages follow $\omega_{it} \overset{ind}{\sim} N_{(0,100)} \left(\omega_t, \tau_{i}^2  \right)$, where $\omega_t$ denotes a statewide average for that variable in year $t$. This community-level process model for the latent variables permits some borrowing of strength across the counties, improving estimation in the counties that only have $5$-year estimates available. We note that a more complicated spatio-temporal structure could be considered here as was done in \citet{Bradley2015}, but it comes with additional computational expense. 

\subsection{Prior Model and Posterior Distribution}

All intercepts and regression coefficients, $\mu_t^{(k)}, \bfbeta^{(k)}, \beta_0^{\mu}, \beta_1^{\mu},$ and $\boldsymbol{\gamma}$, are independently assigned flat, uniform prior distributions on the real line. The statewide average yearly percentages for the ACS variables, $\omega_t$ are assigned a uniform prior distribution over $(0, 100)$. All variance parameters $\sigma^2_k,  \tau^2_k,\tau^2_u,$  $\sigma^2_v$, and $\{\tau^2_i\}$ are assumed to have inverse gamma prior distributions with shape and scale parameters of 0.5. The temporal autoregressive parameters $\phi_k, \phi_u$ are assumed to be uniform over $(0, 1)$. The posterior distribution of the latent processes and parameters is simulated using an adaptive Metropolis-within-Gibbs Markov chain Monte Carlo (MCMC) algorithm implemented using the R package NIMBLE \citep{nimble}. To improve the efficiency associated with sampling highly correlated variables, for each county $i$, the latent counts $N_{i1},...,N_{iT}$ are updated jointly using an automated factor slice sampler as in \citet{Stoner2019}. NIMBLE enforces the zero mean constraint in the ICAR models with a commonly used approach of updating these variates without the constraint and then centering \citep{Paciorek2009}. Convergence of the Markov chain was assessed by visually inspecting trace plots. The R code is included in the supplementary material.

\section{Simulation Study}
\label{se:simulation}

We performed a simulation study to assess the proposed model's ability to accurately predict latent counts of misuse, $N_{it}$, and relative risk $\lambda_{it}$. We fixed values of all hyperparameters and simulated $M=100$ latent processes and data sets from the proposed model where each data set consisted of $T=10$ years of data for each of $n=100$ counties on a $10 \times 10$ grid for $K=2$ county level outcomes and also yearly survey estimates of statewide prevalence. This mimics the setting for the Ohio data we consider in Section \ref{se:app} where the county-level outcomes correspond to treatment admission counts ($k=1$) and death counts ($k=2$). Note that for the purpose of the simulation, we assume the survey estimates are yearly and not multi-year averages, and we assume covariates used in the design matrix $\mathbf{W}$ are known so no process-level model is needed for these values. Specific details regarding how the data were simulated are in the Supplementary Material.

For each of the $M=100$ simulated data sets, we fit the proposed model under two scenarios: (1) assuming we observed yearly survey information and (2) assuming we only observed survey information in years $2, 5, 8$, which we will refer to as the \textit{sparse} survey information scenario. These cases reflect a slightly better and slightly worse case than the actual multi-year average survey data that is available for the application. We will compare the results of our proposed joint model to a baseline model that assumes the state-level survey estimate applies to all counties \citep{Rembert2017,Burke2018} (i.e., no spatial heterogeneity such that $\hat{N}_{it} = \hat{\mu}_{t} P_{it}$) and to a model using only a single county-level outcome (e.g. death counts) \citep{Stoner2019}. We used the R package NIMBLE to run a MCMC algorithm for each of the $M$ data sets under each scenario.

To assess performance, we used the posterior mean as our estimate of latent misuse, $\hat{N}_{it}$, for each simulated data set and computed 95\% equal-tail credible intervals. We used several different criterion to assess how well each approach estimated the true latent counts $N_{it}$ and relative risk $\lambda_{it}$. We computed the coverage probabilities (CP) of the credible intervals for the latent counts, the root mean squared error (RMSE) of the counts and of the relative risk, as well as the relative median absolute error (MAE) for the counts. 

\subsection{Simulation Results}

\begin{table}
\caption{\label{tab:both} A comparison of results from the proposed joint model to the baseline model. The first two columns show the mean and median coverage probability (CP) across the 100 simulated data sets for $N_{it}$ (proposed/baseline). The final three columns present the proportion of simulated data sets for which the proposed model results in a smaller error along with the average error across the 100 data sets for each model (proposed/baseline).}
\centering 
\fbox{%
\begin{tabular}{cccccc}
\hline
Survey & Mean CP & Median CP & RMSE ($N_{it}$) & RMSE ($\lambda_{it}$) & Relative MAE ($N_{it}$)  \\ \hline
 \multirow{2}{*}{Yearly} & \multirow{2}{*}{95\%} & \multirow{2}{*}{96\%} & 100/100 & 100/100 & 100/100 \\ 
 & & & 2598/5581 & 0.29/0.64 & 0.16/0.32  \\ \\
 \multirow{2}{*}{Sparse} & \multirow{2}{*}{93\%} & \multirow{2}{*}{95\%} & 100/100 & 100/100 & 100/100 \\ 
 & & & 2890/5647 & 0.29/0.64 & 0.17/0.33 \\ \hline
\end{tabular}}
\end{table}

\begin{figure}
    \centering
    \makebox{\includegraphics[width=\textwidth]{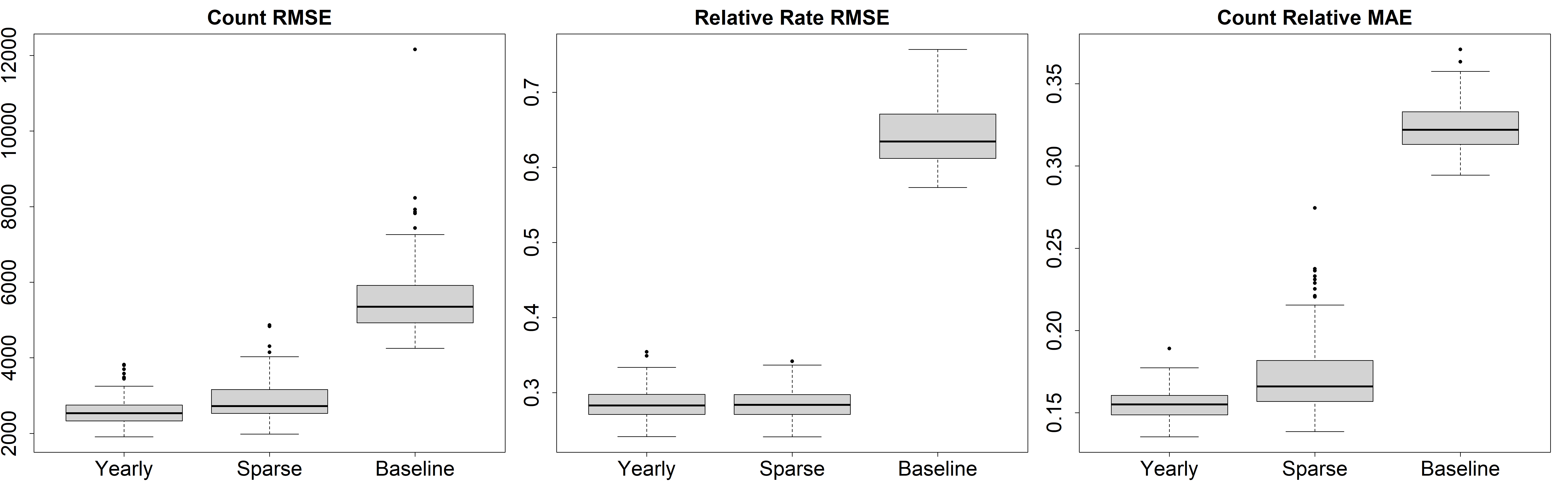}}
    \caption{Boxplots of the RMSE for the counts ($N_{it}$) (left) and relative risk ($\lambda_{it}$) (middle) and the relative MAE for the counts ($N_{it}$) (right) for the 100 simulated data sets under the proposed joint model assuming yearly survey data, the proposed joint model assuming sparse survey data, and the baseline estimates based on yearly survey data.}
    \label{fig:bothoutcomes}
\end{figure}

\begin{figure}
    \centering
    \makebox{\includegraphics[width=\textwidth]{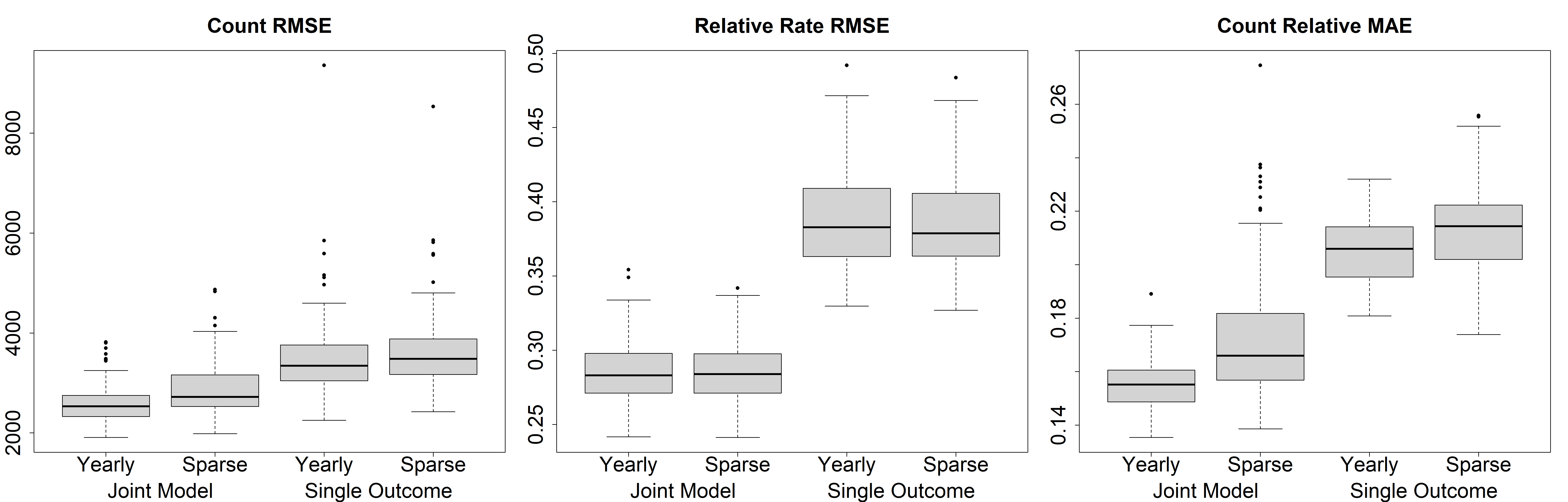}}
    \caption{Boxplots of the RMSE for the counts ($N_{it}$) (left) and relative risk ($\lambda_{it}$) (middle) and the relative MAE for the counts ($N_{it}$) (right) for the 100 simulated data sets under the proposed joint model and the single outcome model for cases assuming yearly and sparse survey data.}
    \label{fig:deathonly}
\end{figure}

The main findings of the simulation study are summarized here with additional results in the supplementary material. Briefly, the error rates of the proposed model were roughly half that of the baseline model for both the yearly and sparse survey scenarios, with these quantities smaller for the yearly survey information setting compared to the sparse survey setting (Table \ref{tab:both} and Figure \ref{fig:bothoutcomes}). The average coverage probabilities of the credible intervals were very close to the target value of 95\%. We also compared the results of our proposed joint model to the model that only considers a single county-level outcome ($Y_{it}^{(2)}$) in addition to the survey data for both the yearly and sparse survey data cases.  The proposed joint model yields a coverage probability that is closer to 95\% than would be obtained modeling just a single county-level outcome. We also see that the proposed joint model yields smaller errors for almost all of the 100 simulated data sets, with the average error reduced by at least 20\% (Table \ref{tab:bothvsdeath} and Figure \ref{fig:deathonly}).

In addition to illustrating that the proposed model has smaller error than competing approaches, we also show that, on average, the proposed model recovers the true parameters of the data generating model. In Supplemental Figure 1, we show that we were able to estimate the intercept parameters $\mu_t$ ($\beta^{\mu}_0$ and $\beta^{\mu}_1$), $\mu_t^{(1)}$, and $\mu_t^{(2)}$. This suggests that we have adequate information on the range of the detection probabilities to overcome the identifiability issues that are common with abundance models. In addition, we show in Supplemental Figure 2 that we are able to estimate the relative risks, $\lambda_{it}$, quite well. Across the 100 simulated data sets, the average difference between the true and estimated $\lambda_{it}$ is -0.003. Likewise, we show scatterplots of the estimates against the true values for three randomly selected simulated data sets in Supplement Figure 3 and see points clustered around the $y=x$ line, which indicates that the model is generally recovering the truth.  Again, this suggests that we are able to recover the true values of the model parameters, suggesting that we have adequate information to inform the model and practically overcome challenges related to identifiability.

In summary, our simulation study showed that our model accurately recovers the latent counts and relative risks and drastically outperforms the baseline model.  We have also shown that our model with sparse survey information still outperforms the baseline model, although it is not as good as having yearly survey data.  We observe that performance is improved by including multiple observed surveillance outcomes compared to using only a single outcome.  Finally, we illustrated that the integration of state-level survey information overcomes limitations of traditional abundance models and adequately informs the estimation of model parameters.

\begin{table}
 \caption{\label{tab:bothvsdeath} A comparison of results from the proposed joint model to the single outcome model. The first two columns show the mean and median coverage probability (CP) across the 100 simulated data sets for $N_{it}$ (proposed/single outcome). The final three columns present the proportion of simulated data sets for which the proposed model results in a smaller error along with the average error across the 100 data sets for each model (proposed/single outcome).}
 \centering
\fbox{%
\begin{tabular}{cccccc}
\hline
Survey& Mean CP & Median CP & RMSE ($N_{it}$) & RMSE ($\lambda_{it}$) & Relative MAE ($N_{it}$)  \\ \hline
 \multirow{2}{*}{Yearly} & \multirow{2}{*}{95\%/89\%} & \multirow{2}{*}{96\%/90\%} & 99/100 & 100/100 & 100/100 \\  
 & & & 2598/3559 & 0.39/0.64 & 0.16/0.21  \\ \\
 \multirow{2}{*}{Sparse} & \multirow{2}{*}{93\%/89\%} & \multirow{2}{*}{95\%/90\%} & 93/100 & 94/100 & 91/100  \\  
 & & & 2890/3665 & 0.39/0.64 & 0.17/0.21 \\ \hline
\end{tabular}}
\end{table}

\section{Ohio Prevalence Estimates}
\label{se:app}

In this section, we apply the model defined in Section \ref{se:data} to estimate the unobserved number of PWMO from 2007-2019 for each of Ohio's $n=88$ counties. To do so, we utilize $T=13$ years of observed county-level counts of opioid overdose death and treatment admissions and multi-year state-level survey estimates of the prevalence of opioid misuse.

\begin{figure}
    \centering
    \makebox{\includegraphics[width=\textwidth]{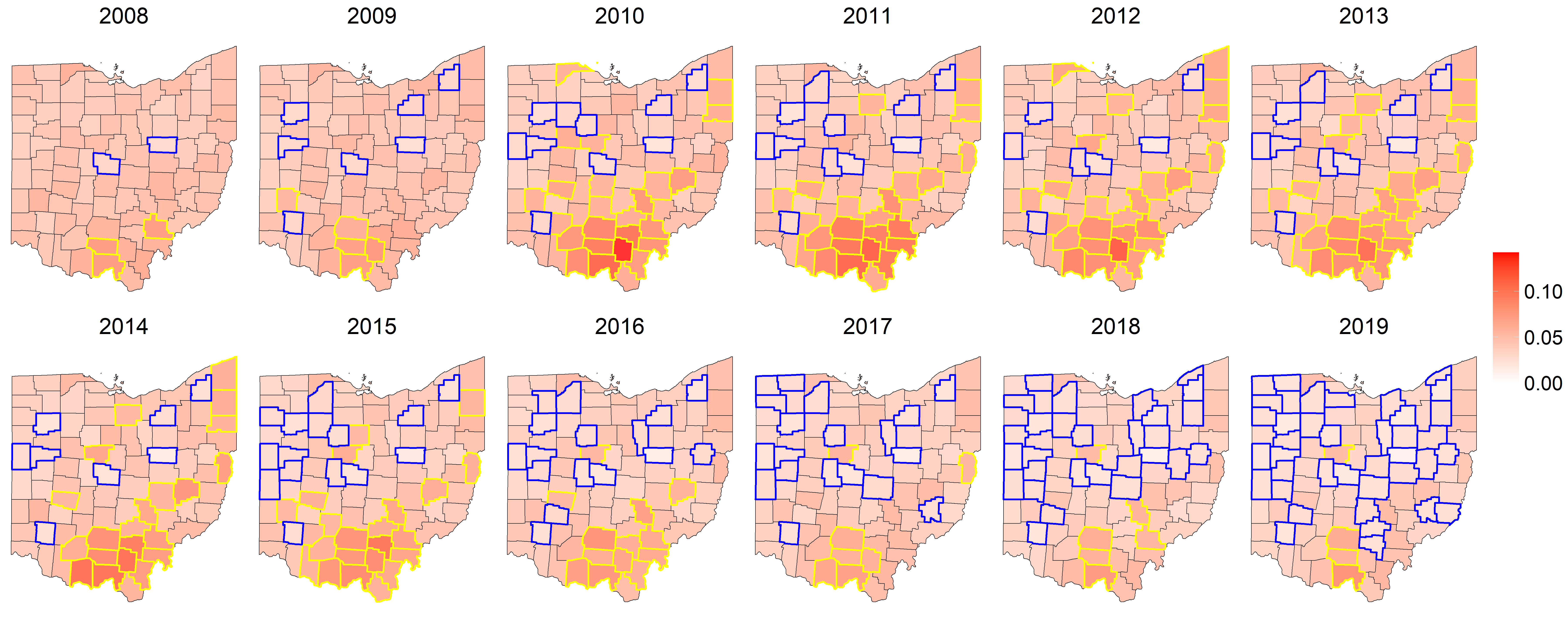}}
    \caption{Maps of the estimated prevalence of PWMO, given by $\hat{N}_{it}/P_{it}$. Counties outlined in yellow have 95\% credible intervals that are entirely above the baseline estimate, and the counties in blue have credible intervals that are entirely below.}
    \label{fig:misuse}
\end{figure}

\begin{figure}
    \centering
    \makebox{\includegraphics[width=\textwidth]{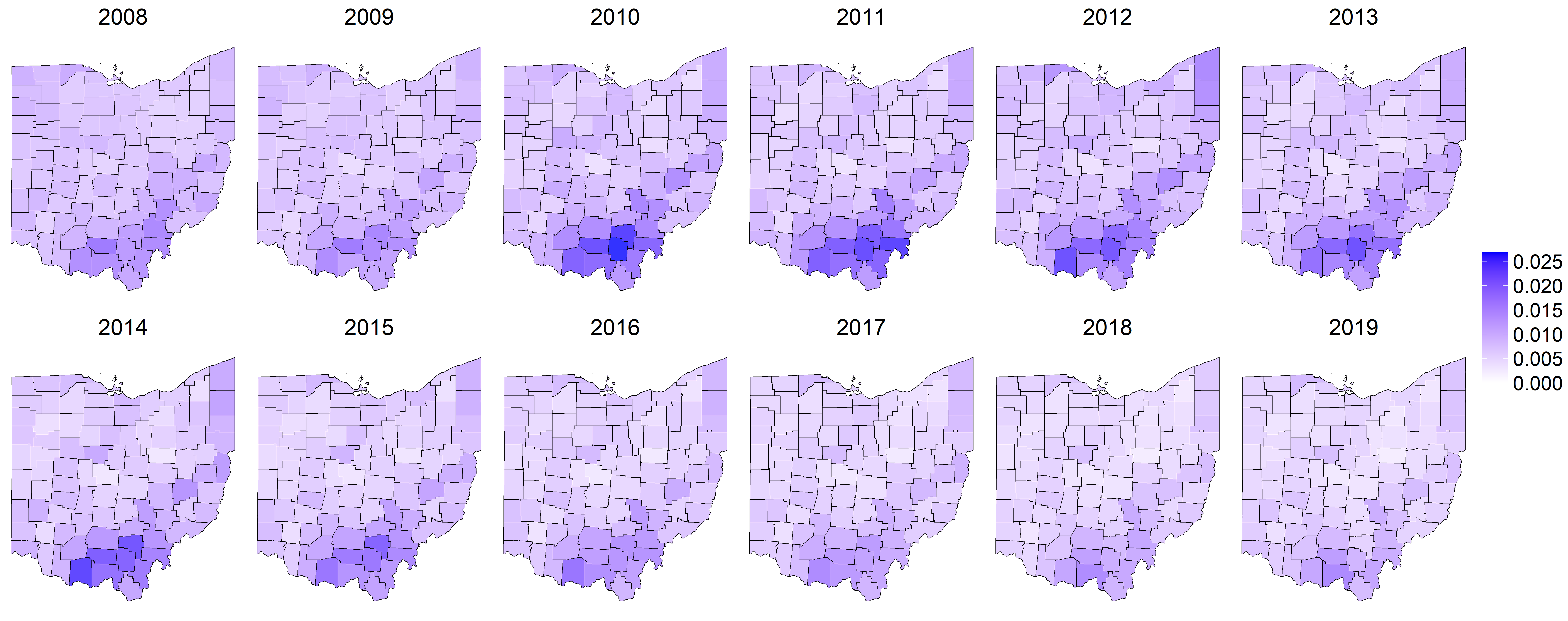}}
    \caption{Maps of the standard error of the prevalence of PWMO.}
    \label{fig:misusese}
\end{figure}

Figure \ref{fig:misuse} maps the posterior mean of the estimate prevalence, $\hat{N}_{it}/P_{it}$, and Figure \ref{fig:misusese} contains the posterior standard deviation. Maps for the estimated log relative risk, $\log\left(\hat{\lambda}_{it}\right)$, are in the supplementary material. We observe county-level and yearly heterogeneity with prevalence ranging from approximately 0 to nearly 0.13. We observe the highest prevalence in southern Ohio, which is commonly considered the epicenter of the opioid epidemic in Ohio.  This map also identifies the counties whose estimated counts of the PWMO are significantly different than would be obtained under the naive baseline estimate that assumes a homogeneous statewide rate, estimated from the survey data. More specifically, the counties outlined in yellow have 95\% credible intervals (CI) that are entirely above the baseline estimate, and the counties in blue have CIs that are entirely below.



The statewide average prevalence of misuse parameters were estimated to be $\hat{\beta}_0^{\mu}=0.0535$ with 95\% CI (0.0516 to 0.0557) and $\hat{\beta}_1^{\mu}=-0.0006$ with 95\% CI (-0.0009 to -0.0004). Table \ref{tab:misuse} contains the corresponding information for the prevalence of misuse regression coefficients. We observe a 5\% increase in prevalence per standard deviation increase in unemployment and high school education, a roughly 10\% increase per standard deviation increase in poverty and food stamps, and a 24\% increase per standard deviation increase in prescribing rate.

\begin{table}
     \caption{\label{tab:misuse} Posterior mean  and 95\% credible intervals of prevalence ratios per 1 standard deviation change in each covariate for prevalence of PWMO.}
     \centering
    \fbox{%
    \begin{tabular}{ccc} \hline
       Variable & Estimate  &  95\% CI \\ \hline
       Poverty  &  1.10   & (1.08, 1.13)  \\
       Unemployment & 1.05 & (1.01, 1.08) \\
       High School & 1.05 & (1.02, 1.08) \\
       Food Stamps & 1.11 & (1.08, 1.15) \\
       Prescribing Rate & 1.24 & (1.19, 1.29) \\ \hline
    \end{tabular}}
\end{table}

\begin{figure}
    \centering
    \makebox{\includegraphics[width=\textwidth]{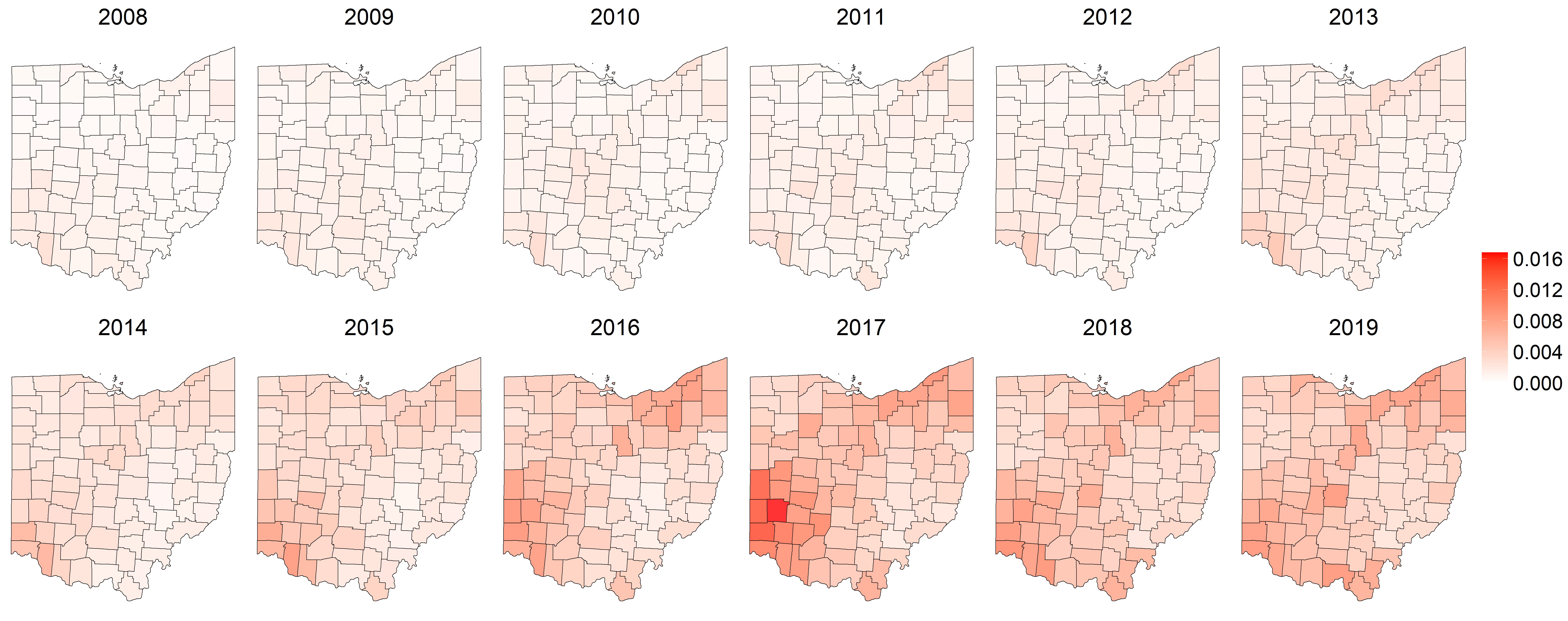}}
    \caption{Maps of the estimated death rate among PWMO, $\hat{p}^{(D)}_{it}$.}
    \label{fig:death}
\end{figure}

Figure \ref{fig:death} is a map of the estimated death rate among PWMO, $\hat{p}^{(D)}_{it}$. A map of the standard errors is in the supplementary material. In the earlier years, we see very low death rates with a large degree of spatial heterogeneity. Starting around 2012, we see increases in the death rate for the southwestern Ohio region corresponding to the Cincinnati, Ohio area, followed by increasing rates in the northeastern Cleveland, Ohio region. This is likely due to the influx of fentanyl in these regions during this time period \citep{Daniulaityte2017,Pardo2019}. Table \ref{tab:deathtrtest} shows the posterior means and 95\% credible intervals for the odds ratios for the death rate. We estimate that the odds of death are 19\% higher in metropolitan statistical areas compared to non-metropolitan areas.

\begin{table}
    \caption{\label{tab:deathtrtest} Posterior means and 95\% credible intervals for the odds ratios corresponding to the covariates in the models for death rate and treatment rate among PWMO.}
    \centering
    \fbox{%
    \begin{tabular}{cccc}
      & Variable & Estimate  &  CI \\ \hline
    \multirow{3}{*}{Death Rate}  & Interstates &    1.04& (0.96,  1.14)  \\
      & HIDTA & 1.07 & (0.97,  1.17) \\
      & MSA & 1.19 & (1.08, 1.30) \\ \hline
      \multirow{2}{*}{Treatment Rate}  & HPSA  &  0.91   & (0.85, 0.98)   \\
       & MU & 1.01 & (0.93, 1.09) \\
      \hline
    \end{tabular}}
\end{table}

\begin{figure}
    \centering
    \makebox{\includegraphics[width=\textwidth]{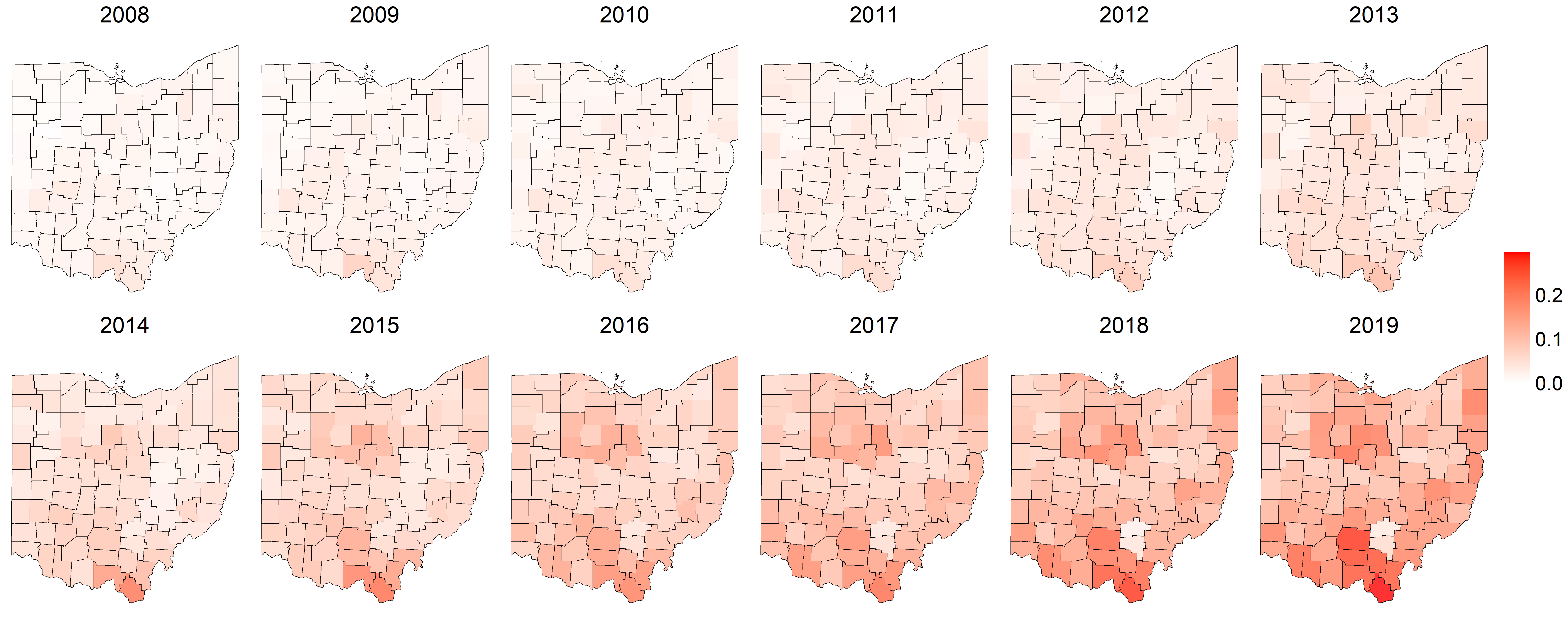}}
    \caption{Maps of the estimated treatment rates $\hat{p}^{(T)}_{it}$ for each county from 2007 to 2018.}
    \label{fig:trt}
\end{figure}

Similarly, Figure \ref{fig:trt} maps the estimated treatment rates among PWMO, $\hat{p}^{(T)}_{it}$. A map of the standard errors is in the supplementary material. We generally see an increasing treatment rate over time, with the largest rates in southern Ohio which is known to have received a large number of resources towards treatment of opioid misuse \citep{Governor2012}. Table \ref{tab:deathtrtest} shows the posterior means and 95\% credible intervals for the treatment rate odds ratios. We estimate that the odds of treatment are 9\% less in health professional shortage areas compared to non-shortage areas.

\begin{figure}[h]
    \centering
    \makebox{\includegraphics[width=.8\textwidth]{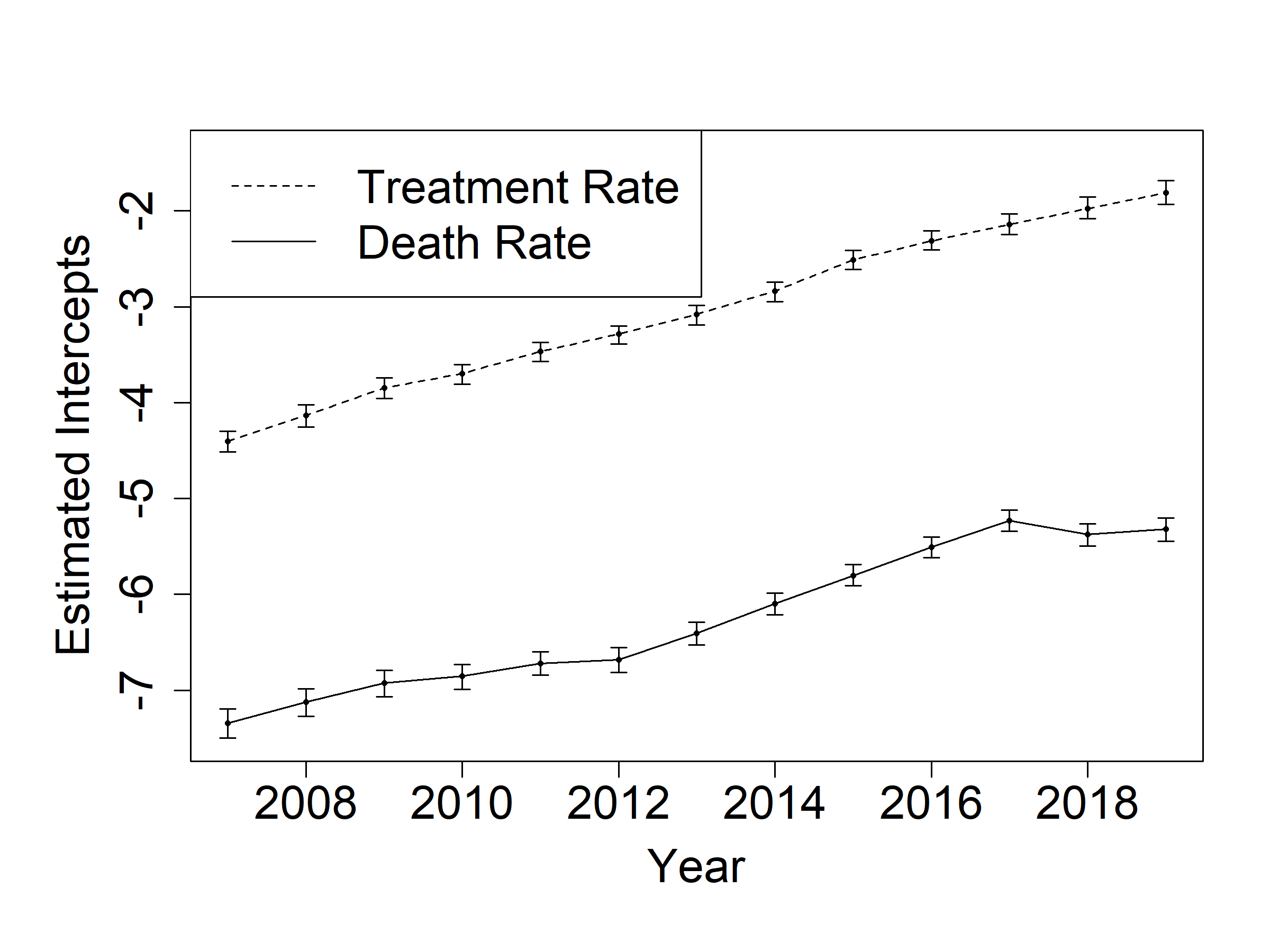}}
    \caption{Plot of the time-varying intercepts for treatment and death}
    \label{fig:trtdeathtrends}
\end{figure}

Figure \ref{fig:trtdeathtrends} plots the time-varying intercepts for death and treatment rates, $\mu_t^{(D)}$ and $\mu_t^{(T)}$. Recall the covariates included for both logistic regression models are indicator variables, so the intercepts for death are interpreted as the yearly average death rates (on the logit scale) among counties without interstates that are neither high impact drug trafficking areas nor metropolitan statistical areas. We generally observe an increasing trend, with a sharp increase beginning in 2012. The years following 2012 correspond to the time in which fentanyl began to infiltrate the state, resulting in a drastic increase in overdose deaths \citep{Pardo2019}. The estimates of $\mu_t^{(T)}$ represent the yearly average treatment rates (on the logit scale) among counties that are neither health professional shortage areas nor medically underserved areas. We see a generally increasing trend with slight shifts observed in 2010 and also in 2014. We note that these time periods align with the passing of state legislation expanding access to treatment \citep{Governor2012}  and also with Medicaid expansion which occurred in Ohio in 2014. 

\section{Discussion}
\label{se:discuss}

In this paper, we developed an approach for estimating the county-level prevalence of PWMO using indirect information from observed, aggregate surveillance outcomes synthesized within an abundance model framework.  By integrating state-level external information, we showed that our model can identify the intercepts at both the level of the latent counts and for the observed surveillance data.  We also show that this approach is superior to assuming homogeneity across counties and to using a single surveillance outcome. By coherently leveraging joint information across data sources, we were able to recover model-based estimates of the latent counts of interest.

We applied our framework to estimate annual county-level counts and prevalence of PWMO in Ohio over a 13 year period.  By doing so, we were able to estimate the prevalence of PWMO, which is extremely relevant for public health policy and until this point, estimates did not exist at the county-level across the state. In addition, we estimated death and treatment rates within the population of PWMO.  This is in contrast to typical epidemiological analyses that define the population at risk as the whole population rather than just the PWMO.  Therefore, the estimates of these rates are more relevant for describing trends in PWMO and informing the targeting of resources and harm reduction interventions. These estimates can also be used to fill data gaps by informing key model parameters in simulation models developed to inform policy choices \citep{Jalali2020}.  We also described associations with covariates at each level of the model.

Our work forms a foundation for this line of research as there are additional methodological challenges to address. For instance, the primary goal of our simulation study was to establish the model's ability to accurately estimate the latent counts and prevalence of PWMO. We have not thoroughly assessed the conditions required to accurately estimate covariate effects. This issue of identifying covariate effects for both observed and latent processes has been studied under various settings (e.g. \cite{Lele2012, Hepler2018Ecology, Stoner2019}), but the results of those papers may not hold since we integrate multiple observed variables at the desired spatial support with additional information at the state level. A future research question is to investigate how well and under what conditions our model can estimate covariate effects. Additional avenues for future research include studying advantages and disadvantages of including more outcome variables and also utilizing our model to evaluate policy and optimize resource allocation.

Our analysis has several limitations.  First, we assume the surveillance outcomes are observed without error, but there is potential for misclassification, particularly of overdose deaths \citep{Slavova2015}.  However, Ohio is considered to have excellent reporting of overdose deaths \citep{Scholl2019}.  We also use survey estimates to inform the intercepts which are potentially underreported.  While the model can flexibly adjust estimates around the parameters informed by the survey, future work will formally explore sensitivity to bias in the survey estimates. In addition, the language of the survey question addressing opioid misuse was changed in 2015 which may have impacted responses.  We also assume that all individuals counted as a treatment admission or an overdose death belong to the population of PWMO.  While this is a reasonable assumption, it is unlikely to be universally true, particularly as fentanyl is unknowingly added to other substances \citep{Mars2019,Townsend2021}. In addition, all analyses are aggregate and should be interpreted at the appropriate level to avoid the ecological fallacy \citep{Piantadosi1988}.

In conclusion, we have a developed a model within the abundance model framework to estimate the size of hidden populations using observed data that provide indirect information.  Through synthesis of multiple sources of data, we are able to generate model-based estimates of hidden quantities that are critical for informing public health policy and the allocation of resources.  We believe this is a promising framework for addressing questions about hidden epidemiological populations and can provide a foundation for future research.

\section*{Acknowledgements}

Research reported in this publication was supported by the National Institute On Drug Abuse of the National Institutes of Health under Award Number R21DA045236 and the National Institute of Child Health and Human Development under Award Number R01HD092580. The content is solely the responsibility of the authors and does not necessarily represent the official views of the National Institutes of Health. 

These data were provided by the Ohio Department of Health. The Department specifically disclaims responsibility for any analyses, interpretations or conclusions.

\bibliographystyle{rss}
\bibliography{r21_refs}

\newpage

\begin{center}
	\LARGE{Supplemental Material}
\end{center}

\setcounter{figure}{0}
\setcounter{table}{0}

\begin{figure}[h]
	\centering
	\subfigure[$\beta_0^{\mu}$]{\includegraphics[width=.47\textwidth]{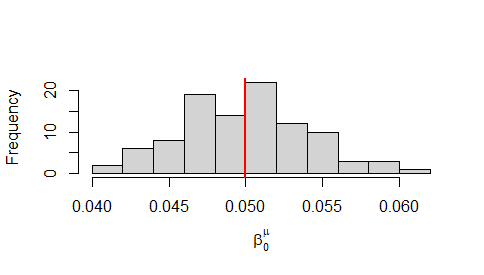}}
	\subfigure[$\beta_1^{\mu}$]{\includegraphics[width=.47\textwidth]{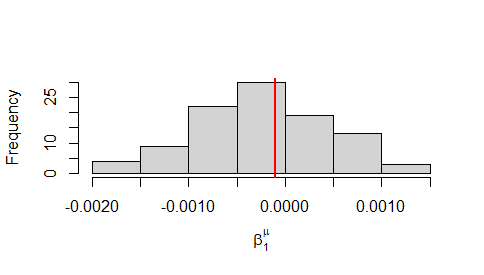}}
	\subfigure[$\mu_t^{(1)}$]{\includegraphics[width=.47\textwidth]{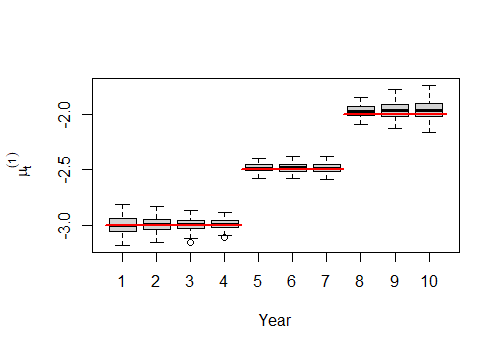}}
	\subfigure[$\mu_t^{(2)}$]{\includegraphics[width=.47\textwidth]{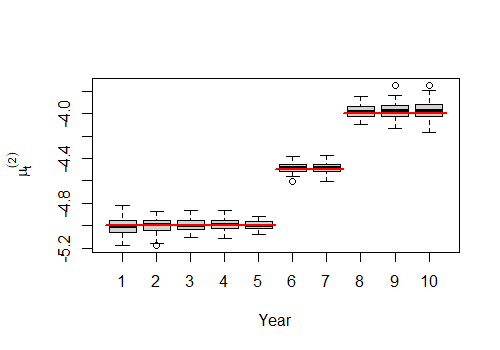}}
	\caption{Plots of the estimates of the intercept parameters for the $H=100$ simulated data sets. The vertical red lines correspond to the true values used to simulate the data.}
\end{figure}

\begin{figure}
	\centering
	\subfigure[Absolute Difference]{\includegraphics[height=.5\textheight]{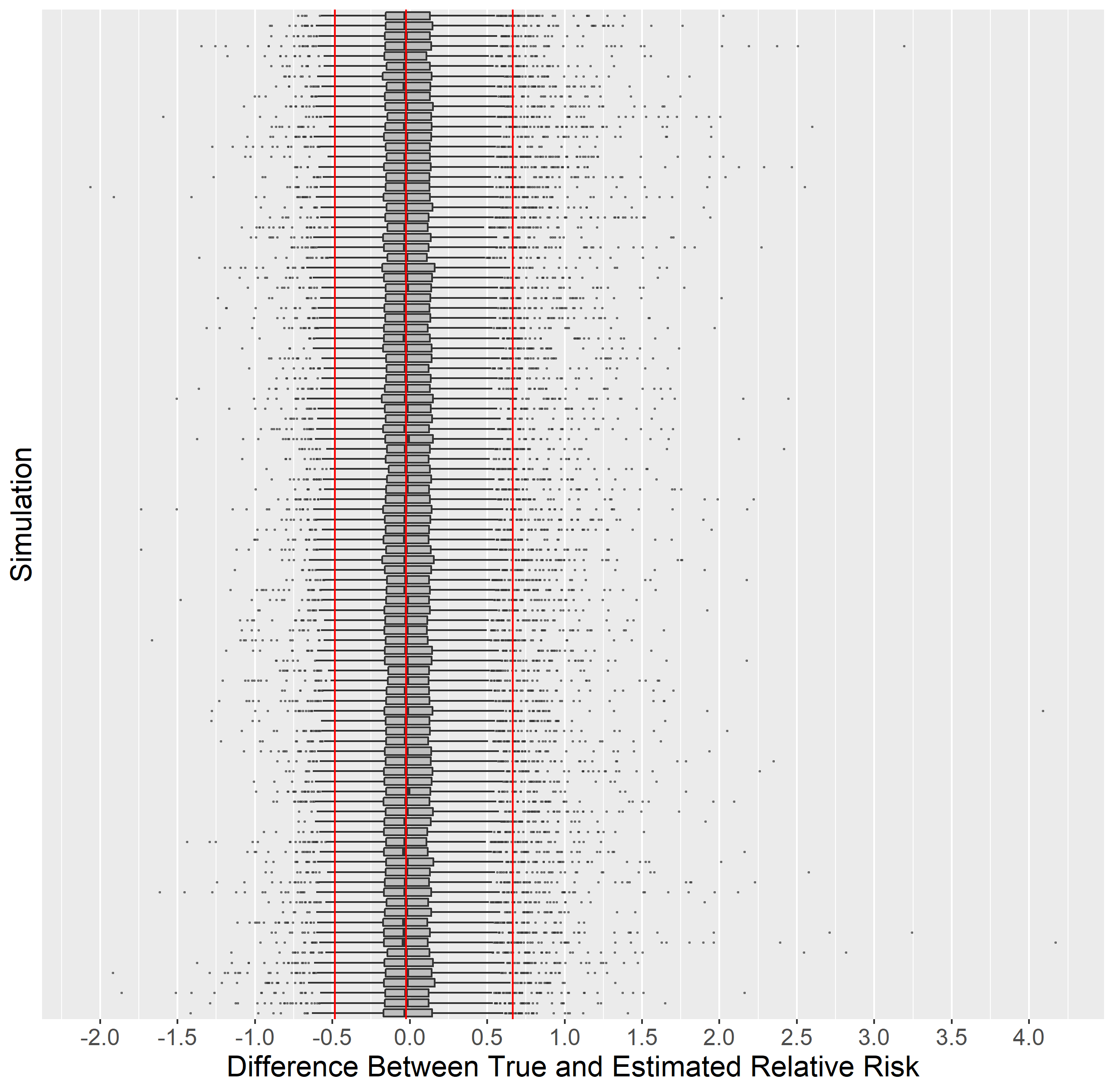}}
	\subfigure[Relative Difference]{\includegraphics[height=.5\textheight]{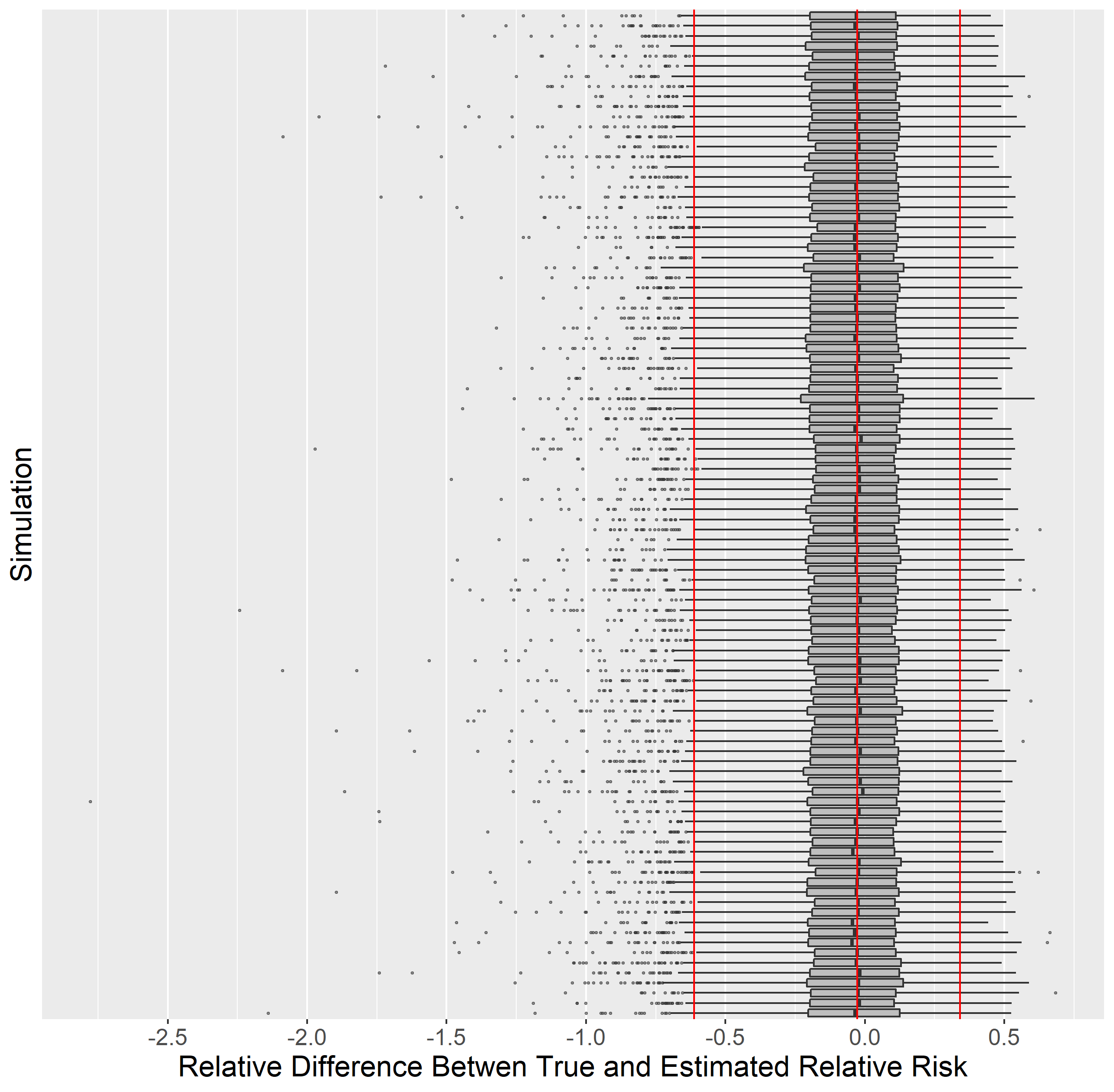}}
	\caption{Boxplots of the absolute and relative difference between the estimate of $\lambda_{it}$ and the true simulated value for each location in each of the $H=100$ simulated data sets. Vertical red lines correspond to the first quartile, median, and third quartile across all of the simulated data sets.}
\end{figure}

\begin{figure}
	\centering
	\subfigure[$N_{it}$]{\includegraphics[height=.42\textheight]{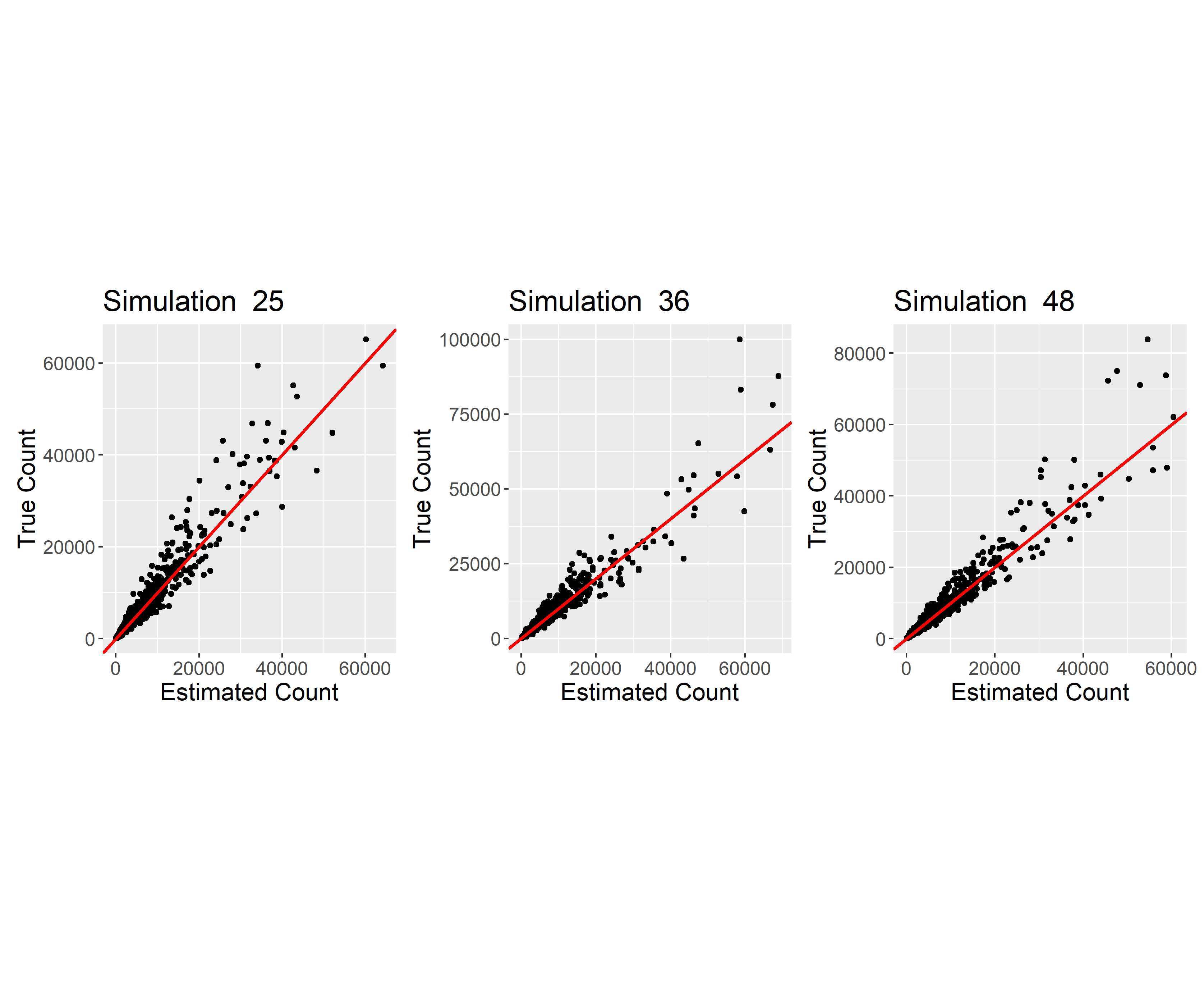}}
	\subfigure[$\lambda_{it}$]{\includegraphics[height=.42\textheight]{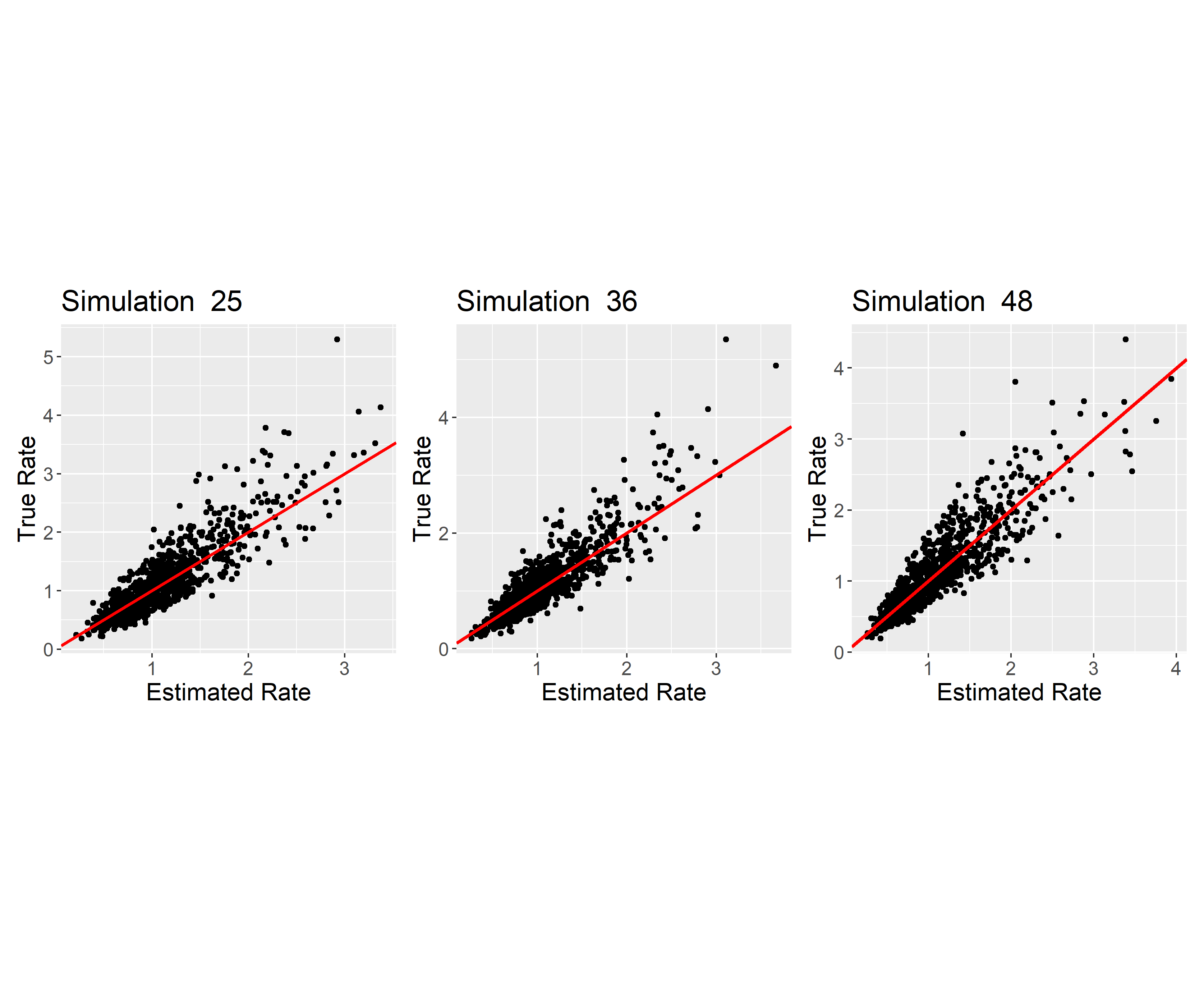}}
	\caption{Scatterplots from randomly selected simulated data sets showing the estimated and true counts ($N_{it}$) and relative risks ($\lambda_{it}$). Red lines indicate when the estimates equal the truth.}
\end{figure}

\begin{table}
	\caption[caption]{Estimates of the average statewide prevalence of misuse \\\hspace{\textwidth}over the given time frame along with standard errors of the estimates.}
	\fbox{
		\begin{tabular}{ccc}
			Years & Estimate & Standard Error \\ \hline
			2003 - 2006 & 0.05 & 0.0025 \\
			2007 - 2010 & 0.055 & 0.0026 \\
			2011 - 2014 & 0.052 & 0.0024 \\
			2015 - 2016 & 0.047 & 0.0041 \\
			2016 - 2017 & 0.051 & 0.0037 \\
			2017 - 2018 & 0.041 & 0.0033 \\
			2018 - 2019 & 0.043 & 0.0043
	\end{tabular} }
\end{table}

\begin{figure}
	\centering
	\makebox{\includegraphics[width=\textwidth]{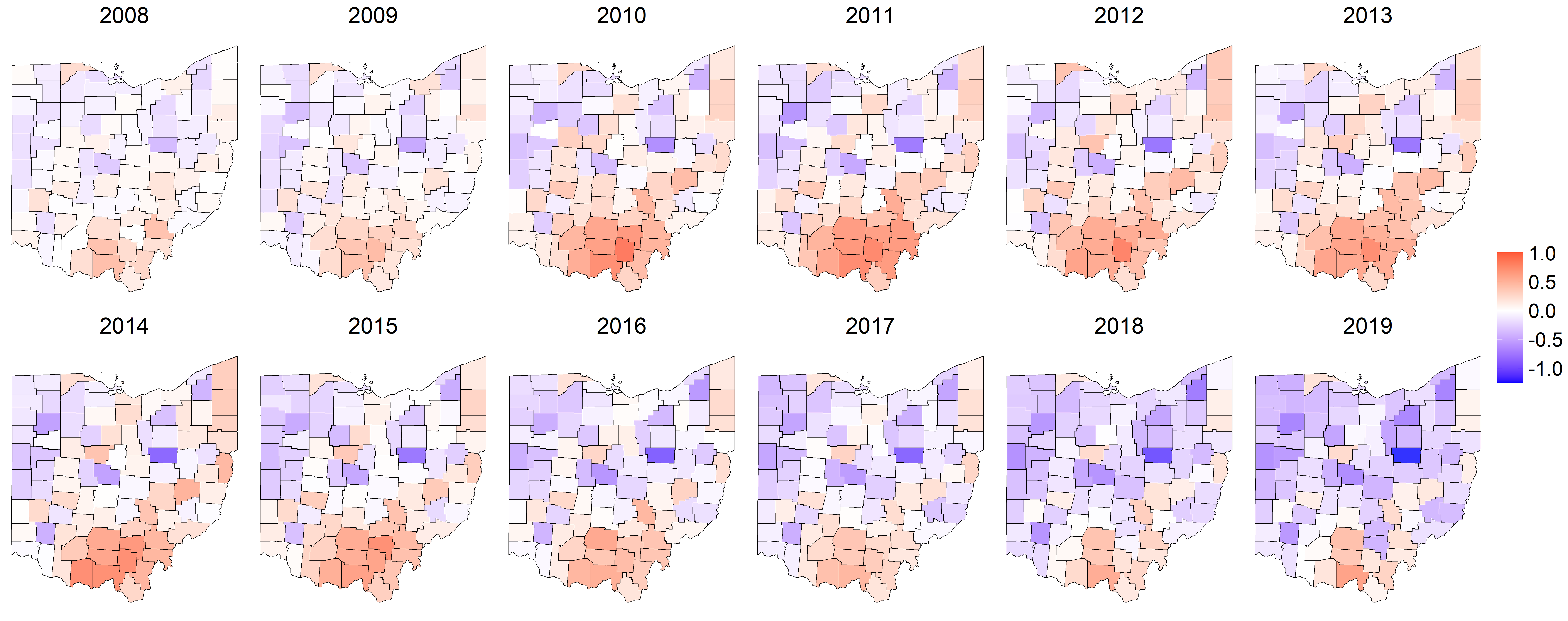}}
	\caption{Maps of the estimated relative risk of PWMO, given by $\lambda_{it}$.}
\end{figure}

\begin{figure}
	\centering
	\makebox{\includegraphics[width=\textwidth]{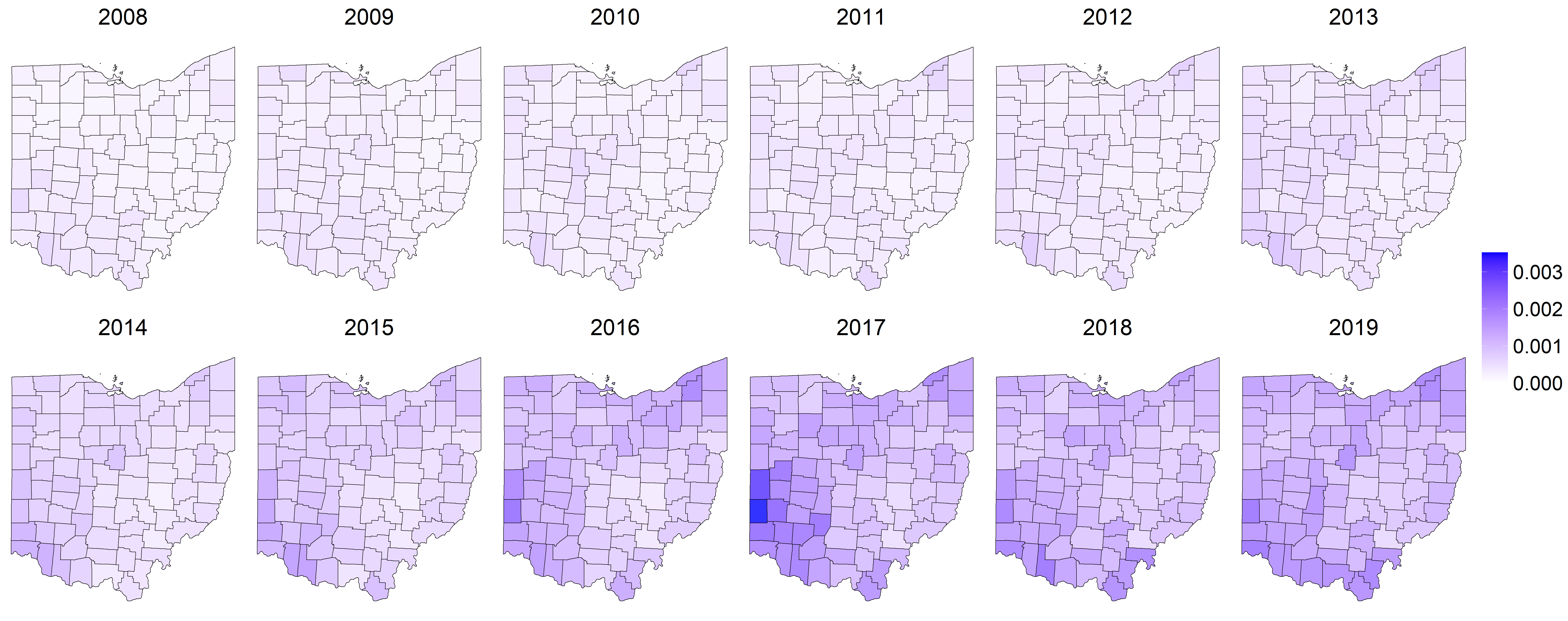}}
	\caption{Maps of the posterior standard deviation of the death rate, $p^{(D)}_{it}$.}
\end{figure}

\begin{figure}
	\centering
	\makebox{\includegraphics[width=\textwidth]{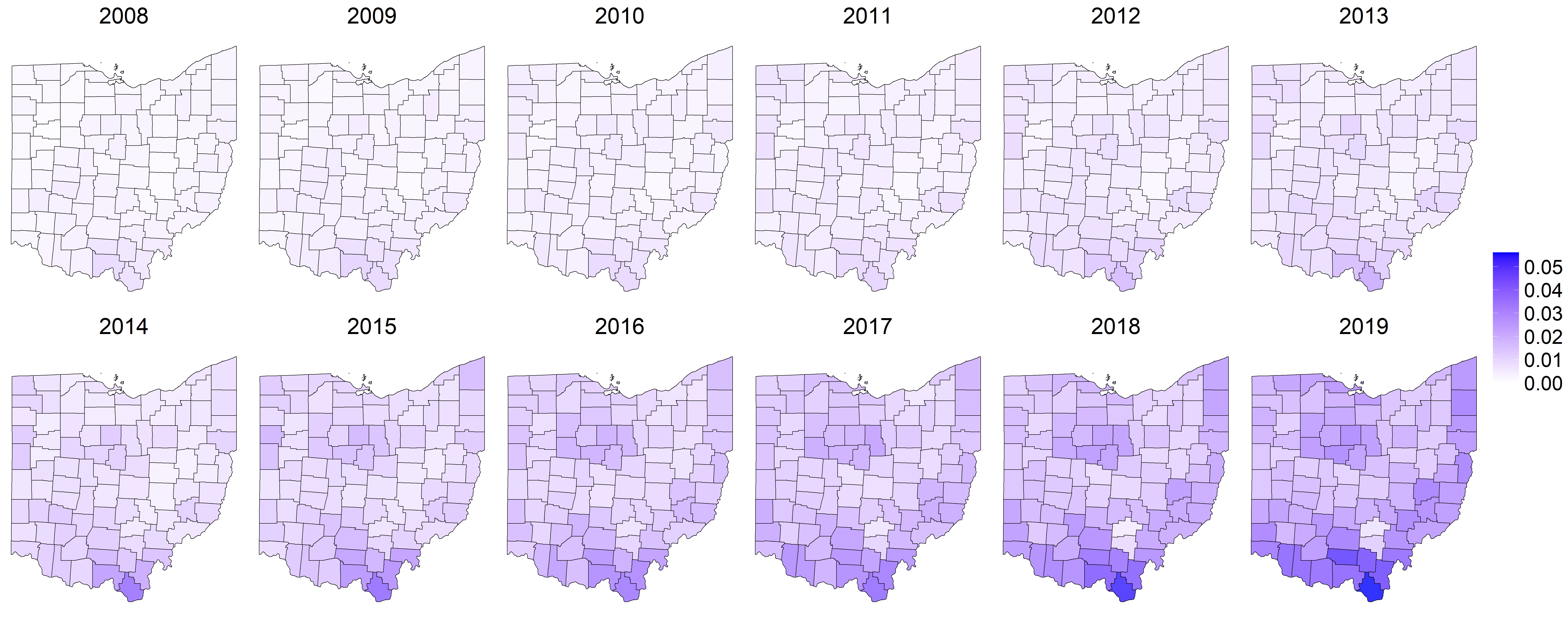}}
	\caption{Maps of the posterior standard deviation of the treatment rate, $p^{(T)}_{it}$.}
\end{figure}

\end{document}